\documentclass[superscriptaddress,footnoteinbib,showpacs,amssymb,twocolumn,aps]{revtex4-1}

\usepackage{graphicx,amsmath,color}
\usepackage{booktabs,array,dcolumn}

\newcommand{\fig}[1]{Fig.~\ref{#1}}
\newcommand{\eq}[1]{Eq.~(\ref{#1})}
\newcommand{\bn}{ {\bf \hat{n}} }
\newcommand{\be}{ {\bf \hat{e}} }

\newcommand{\bx}{ {\bf \hat{x}} }
\newcommand{\by}{ {\bf \hat{y}} }
\newcommand{\bz}{ {\bf \hat{z}} }
\newcommand{\bw}{ {\bf \hat{w}} }
\newcommand{\bv}{ {\bf \hat{v}} }
\newcommand{\bhu}{ {\bf \hat{u}} }

\newcommand{\br}{ {\bf r} }
\newcommand{\pp}{{\mathcal P}_{2}}
\newcommand{\eeq}{ \end{equation} }
\newcommand{\beq}{ \begin{equation} }
\newcommand{\eea}{ \end{eqnarray} }
\newcommand{\bea}{ \begin{eqnarray} }
\newcommand{\p}{ ^{\prime}}

\begin{document}

\title{Frank elasticity of composite colloidal nematics with anti-nematic order}
\author{H. H. Wensink }
\affiliation{Laboratoire de Physique des Solides, CNRS, Universit\'e Paris-Sud, Universit\'e Paris-Saclay, 91405 Orsay, France}
\email{rik.wensink@u-psud.fr}

\date{\today}

\begin{abstract}
Mixing colloid shapes with distinctly different anisotropy generates composite nematics  in which the order of the individual components can be fundamentally different. In colloidal rod-disk mixtures or hybrid nematics composed of anisotropic colloids immersed in a thermotropic liquid crystal, one of the components may adopt so-called anti-nematic order while the other exhibits  conventional nematic alignment.  Focussing on simple models for  hard rods and disks, we employ Onsager-Straley's second-virial theory to derive scaling expressions for the elastic moduli of rods and disks in both nematic and anti-nematic configurations and identify their explicit dependence on particle concentration and shape. We demonstrate that the splay, bend and twist elasticity of anti-nematically ordered particles scale logarithmically with the degree of anti-nematic order, with the bend-splay ratio for anti-nematic discotic nematics being far greater than for conventional nematic systems.  The impact of surface anchoring on the elastic properties of hybrid nematics will also be discussed in detail.  We further demonstrate that the elasticity of mixed uniaxial rod-disk nematics depends exquisitely on the shape of the components and we provide simple scaling expressions that could help engineer the elastic properties of composite nematic liquid crystals. 

\end{abstract}

\maketitle

\section{Introduction} 

Understanding the structure-property relationship of nematic liquid crystals -- fluids without long-ranged positional order composed of aligned non-isotropic molecules or nanoparticles transmitting long-ranged orientational order  -- relies strongly on knowledge of the nemato-elastic properties and their dependence on thermodynamic properties such as temperature or density \cite{gennes-prost}.  For a bulk nematic fluid  in three spatial dimensions there are essentially three fundamental modes, namely splay, twist and bend,  that describe the extent of thermally driven elastic deformations of the uniform director field \cite{frank}.   In liquid-crystal based devices, the electro-optic switching characteristics of the director field are largely controlled by the splay and bend elastic modes, while the twist mode plays a key role in determining the helical pitch of chiral liquid crystals \cite{schadt1997,hiltrop}.

Connecting the amplitude of the various elastic modes  to the microscopic properties of the constituents remains a challenging problem in view of the orientation-dependent interactions between the molecules as well as the various attractive and repulsive forces acting among them whose range and amplitude are often unknown. In common modelling practice, one usually resorts to simple coarse-grained potentials or emblematic particle shapes (e.g. a cylindrical rod, disk or ellipsoid) to arrive at a satisfactory microscopic description of the material, including its elastic properties \cite{mederos2014,obrien2011,wilson2005,care2005}.   For lyotropic systems consisting of rigid or semi-flexible nanoparticles in a structureless solvent,  the role of the (effective) particle shape is routinely expressed in terms of the aspect ratio of the components which, along with the particle concentration, constitutes the main parameter controlling the liquid crystal structure and phase behavior  \cite{Onsager49,vroege92, mcgrother1996, bolhuis1997}.  

Obtaining first-principle predictions of the elastic properties from theory or simulation  for elongated colloidal particles is a challenging task in view of the mesoscopic long-wavelength character of director fluctuations which necessitates extensive simulation setups and considerable numerical effort \cite{allen1988a,allen1988a_err,cleaver1991,allenevans,wilson2005,humpert2015a}.  Moreover, strongly anisotropic cylindrical objects suffer from poor sampling statistics in Monte Carlo methods rendering equilibrium structure investigations numerically cumbersome \cite{allenevans, bolhuis1997}.

Density functional theoretical approaches for common nematics of cylindrical mesogens  have been applied quite fruitfully to quantifying the elastic moduli as a function of particle anisotropy,  concentration or  nematic order parameter \cite{odijkelastic,vroege1987,wittmann2015b,anda2018}. However,  extending these theoretical treatments to analyze composite nematics comprising several components of different size and shape has been mostly overlooked thus far. Naively, one could argue that the elasticity of these multicomponent nematics should simple follow from a linear superposition of the elastic contributions of the individual components, weighted by their respective mole fraction. While this approach could certainly be a plausible one for nematic compounds in which all components obey the same nematic symmetry, it is no longer a viable route to describing multi-component nemato-elasticity when different nematic symmetries are present.  For instance, particles may,  under certain circumstances,  generate {\em anti-nematic} orientational order  characterized by a uniaxial symmetry of particles aligned across a plane perpendicular to the nematic director \cite{sokalski1982}. This gives rise to a {\em negative} nematic order parameter, contrary to what is observed in most liquid crystals.

\begin{figure}
\begin{center}
\includegraphics[width= \columnwidth]{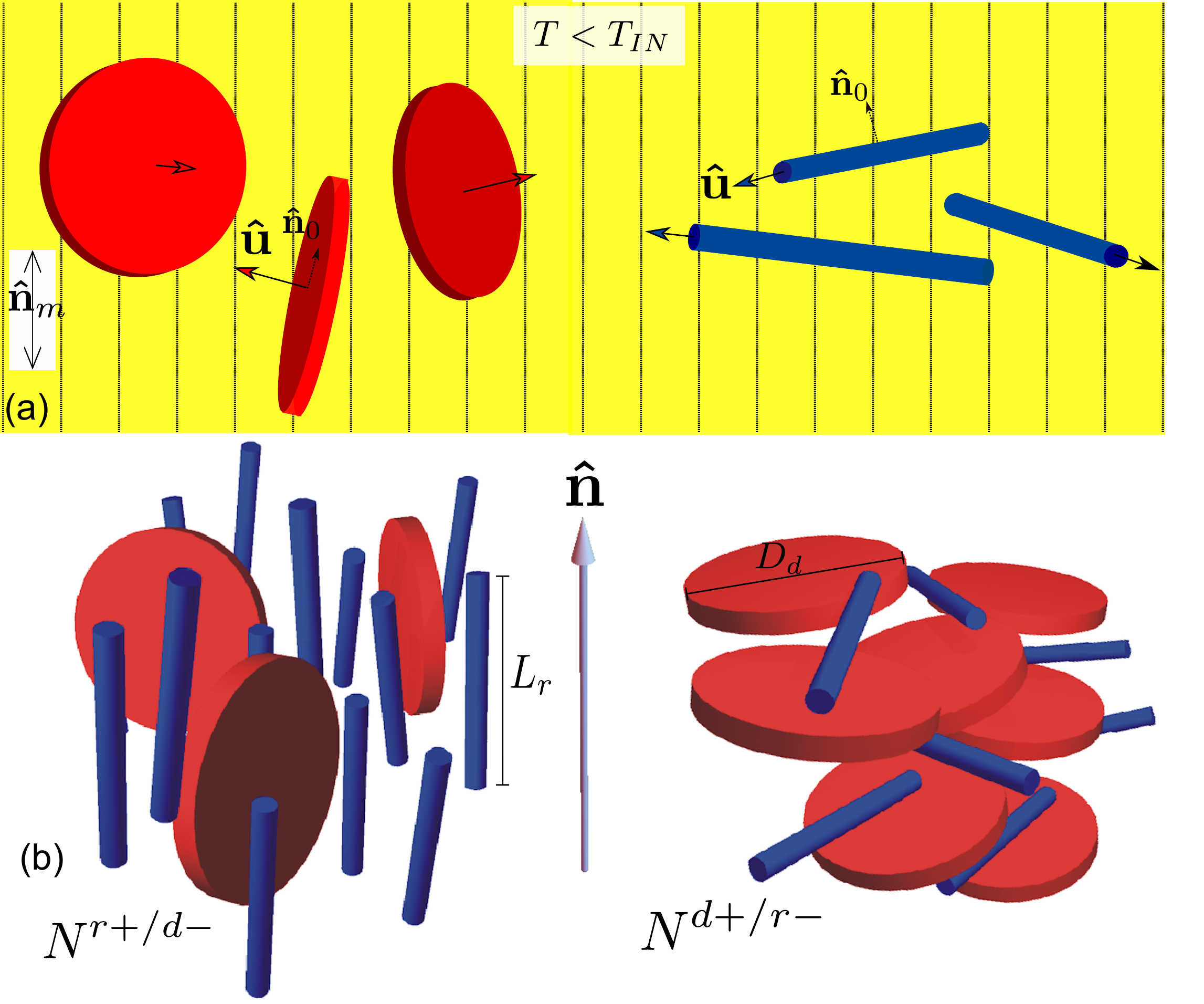}
\caption{ \label{fig1} (a) Anti-nematic order in hybrid nematic materials  generated through surface-anchoring of colloidal inclusions (thin rods or disks) immersed in a thermotropic nematic liquid crystal with molecular director $\bn_{m}$.  (b) Anti-nematic order in mixed colloidal nematics composed of rod- and disk-shaped particles. In the rod-dominated phase (left) the disk normals are aligned in an anti-nematic configuration with their normals pointing perpendicular to the overall director $\bn$. In the disk-dominated phase (right) the rod-shaped inclusions are aligned anti-nematically. }
\end{center}
\end{figure}

Examples where such order may occur are so-called hybrid colloidal-molecular liquid crystals (\fig{fig1}a) or nematics of mixed mineral colloids involving a component with a distinct anti-nematic orientational organization \cite{Liu2016,Mundoor18,smalyukh2018}. In  hybrid nematic systems, small concentrations of strongly anisotropic colloidal particles are immersed in a thermotropic liquid crystal and spontaneously adopt anti-nematic order driven by an interplay between surface-anchoring  and elasticity-mediated interactions enforcing the colloids to align orthogonal to the molecular director \cite{smalyukh2018, Mundoor18}. Other examples are mixed nematics composed of purely repulsive rod- and disk-shaped colloids \cite{kooijrodplate,kooij2000,woolston2015}  where the anti-nematic order of one component is generated spontaneously through excluded-volume-minimizing steric interaction with the other component (see \fig{fig1}b). 
For single component systems,  anti-nematic order has been observed in some soft-interaction models for clay particles \cite{jabbari2014}, deformable dendrimers \cite{georgiou2014}, clay platelets  exposed to an external electric-field  \cite{dozov2011} and lacuna smectic phases composed of ring-shaped mesogens \cite{avendano2009,avendano2016}.

An important condition for stable anti-nematicity in the composite nematics mentioned above is that the mole fraction of the anti-nematic component should be low enough to ensure uniaxial order to be preferable over {\em biaxial} order in which both species are oriented along mutually orthogonal directors, thus imparting an orthorhombic symmetry onto the fluid \cite{vroege2014,Mundoor18}.     
 
In an effort to further our understanding of these composite nematic materials we  wish to address the question as to how nemato-elasticity could be quantified in a multi-component nematic in which one or several components exhibit anti-nematic order. In this work, we use a tractable second-virial theory applied to a representative model system of hard rods and disks to systematically investigate the elastic moduli for both nematic and anti-nematic order. By building on Odijk's asymptotic analysis for the orientation distribution \cite{odijkelastic, odijkoverview} we can work out the different elastic modes in explicit form and gauge their scaling with particle concentration and aspect ratio.  While the results for the rod-  and disk-based  nematics are in qualitative agreement with earlier predictions, we find that the interrelation of the splay, twist and bend modes for the anti-nematic fluid is different from that of a common nematic fluid. Focussing on hybrid nematics, involving strongly anisotropic colloids immersed in a thermotropic nematic background, we argue that surface anchoring has a considerable impact on the elastic properties of the compound nematic material, mostly boosting the splay elasticity in case of needle-type inclusions while for disk-shaped particles the bend mode is significantly enhanced.   We finish our analysis by addressing the nemato-elasticity of rod-disk mixtures where coupling terms describe the elastic response transmitted by steric interactions between the rods and the disks. We propose scaling results that enables one to predict in detail how doping a rod-based nematic with disks (or vice versa) will affect the elastic properties of the reference system. 

The rest of this paper is organized as follows. We begin by describing the principal context of the model and introducing the starting expressions. The subsequent two Sections are devoted to a computation of the scaling expressions for the elastic moduli  of nematic and anti-nematic fluids of disks which we compare to previously derived results for rods. In  Section  V we address the role of colloid surface anchoring on the elastic properties of hybrid molecular-colloidal nematic materials. In Section VI we set out to quantify the effect of interspecies interactions on the elastic properties of nematic phases composed of a rod-disk mixture. Finally, some concluding remarks are formulated in Section VII.

\section{Free energy of director deformation} 

Let us consider a bulk nematic fluid composed of anisotropic building blocks and define elementary director fluctuations of amplitude $\varepsilon  \ll \sigma^{-1}$ (with $\sigma$ the typical particle size) around the nematic director $\bn = \bz$ fixed along the $z-$direction of a Cartesian frame  defined  as ${\bf R} =\{X \bx, Y \by, Z \bz\} $ \cite{allenevans}:
\begin{align}
 \bn_{1}({\bf R} )& = \bn   + \varepsilon Y \by \nonumber \\
 \bn_{2}({\bf R} ) &= \bn   + \varepsilon X \by  \nonumber \\
 \bn_{3} ({\bf R} ) &= \bn   + \varepsilon Z \by,  
 \label{n123}
\end{align}
corresponding to the splay, twist and bend modes, respectively. 

Following Onsager-Straley theory \cite{straley76,allenevans,odijkelastic}, we assume that the free energy due to elastic distortions of the director field  $\bn$ can be quantified by expanding Onsager's original free energy functional \cite{Onsager49} for nematics with a weakly non-uniform director field:
\begin{align}
\delta F_{el} & \sim -\frac{\rho^{2} }{4}\langle \langle  \int d {\bf R}  \int \Delta \br \Phi (\Delta \br;  \bhu ,\bhu \p) (\bar{\nabla}_{\bhu} \bn) (\bar{\nabla}_{\bhu \p} \bn) \rangle \rangle_{\dot{f}} \nonumber \\
& = \frac{1}{2} V K \varepsilon^{2},
\label{fel}
\end{align}
Here, we adopt short-hand notation for the double orientational average $\langle \langle (\cdot) \rangle \rangle_{\dot{f}}  = \int d \bhu \dot{f}(\bhu \cdot \bn ) \int d \bhu \p \dot{f} (\bhu \p \cdot \bn ) (\cdot ) $ and director gradient $\bar{\nabla}_{\bhu} \bn  = (\Delta \br \cdot \nabla) (\bn ({\bf R}) \cdot \bhu)$. For notational compactness, we set the thermal energy $k_{B}T$ to unity without loss of generality. The principal variables in the above expressions are $\dot{f}(x) = \partial f  / \partial x$ the derivative of the uniaxial orientational probability $f(\bhu \cdot \bn)$ for the orientation vector $\bhu$ of each particle, with $V$ is the system volume, $\rho=N/V$ the particle density, and $K$ the elastic modulus.  The key input in the present second-virial approach is the Mayer function  $\Phi = e^{-U} -1$ which contains all the information on the shape and interaction between the constituents through the pair potential $U$ \cite{Onsager49}. In this, study we assume the colloidal interactions to be  hard so that for any particle pair at centre-of-mass distance $\Delta \br$ and orientations $\bhu$ and $\bhu \p$:
\beq
\Phi(\Delta \br ; \bhu, \bhu\p) = \left.
  \begin{cases}
    -1, & \text{if cores overlap} \\
   0, & \text{otherwise } 
      \end{cases}
  \right\} 
\eeq
The principal model we consider is a hard cylinder with aspect ratio $L_{r}/D_{r} \rightarrow \infty$ corresponding to infinitely long hard rods and the inverse limit $L_{d}/D_{d} \rightarrow 0$ describing infinitely thin hard disks. These shapes are considered as emblematic models for a vast range of many lyotropic liquid crystals.  Soft interactions arising from e.g, charge-mediated interactions or any other type of coarse-grained dispersion forces, could, in principle, be included through some effective second-virial theory \cite{francomelgar, wensink_trizac14} but this seriously complicates the present analysis and goes beyond the scope of the current project.

Plugging in the fluctuations $\bn_{i}({\bf R})$ into \eq{fel}  enables us to read off  the microscopic expressions for the respective modes:
\begin{align}
K_{1} & \sim -\frac{\rho^{2}}{2} \langle \langle  \int d \Delta \br  (\Delta y)^{2} \Phi (\Delta \br, \bhu ,\bhu \p) u_{y} u\p_{y}  \rangle \rangle_{\dot{f}} \nonumber \\ 
K_{2} & \sim -\frac{\rho^{2}}{2} \langle \langle  \int d \Delta \br   (\Delta x)^{2} \Phi (\Delta \br, \bhu ,\bhu \p) u_{y} u\p_{y}  \rangle \rangle_{\dot{f}} \nonumber \\ 
K_{3} & \sim -\frac{\rho^{2}}{2} \langle \langle  \int d \Delta \br   (\Delta z)^{2} \Phi (\Delta \br, \bhu ,\bhu \p) u_{y} u\p_{y}  \rangle \rangle_{\dot{f}}, 
\label{k123}
\end{align}
These expressions serve as our starting point for an explicit calculation of the moduli for different particle shapes (which determines the interaction kernel  $\Phi$) and orientational symmetry. The latter is governed by the orientation distribution $f(\bhu \cdot \bn)$) which in turn follows from an optimization of the orientational and excluded-volume entropies contributions to the free energy. This leads to the following  self-consistency expression:  
\begin{align}
f (\bhu \cdot \bn) &= {\mathcal N} \exp \left [  \rho  \langle \int d \Delta \br \Phi (\Delta \br, \bhu ,\bhu \p) \rangle_ {f}   \right ] \nonumber \\ 
&= {\mathcal N} \exp \left [ - \rho  \langle v_{ex}( \bhu ,\bhu \p) \rangle_ {f}   \right ] 
\end{align}
with ${\mathcal N}$ a  normalization factor ensuring $\langle f \rangle_{f}   = 1$.  The Boltzmann probability is thus defined in terms of the ensemble-average of the excluded volume $v_{ex}(\bhu, \bhu \p)$   between anisotropic particles  which itself depends on $f$.   To describe common nematic order, we shall employ a much simpler Gaussian form appropriate for situations where the nematic alignment of the particles is asymptotically strong \cite{odijkoverview}:
\beq
f_{G}  (\theta ) \sim \frac{\alpha}{4 \pi } \exp \left (-\frac{1}{2} \alpha \theta^{2} \right ), \qquad (\alpha \gg 1)
\label{gauss}
\eeq 
in terms of a polar angle $\theta = \cos ^{-1} (\bhu \cdot \bn) $ and  concentration-dependent variational parameter $\alpha$.  Since the orientational order we consider is strictly apolar, the distribution must be supplemented with its mirror form $f_{G}  (\pi  -\theta )$ to reflect the equivalence between alignment along $\bn$ and $-\bn$.  Further on in this study, we shall define the equivalent distribution for the anti-nematic case. Although the Gaussian distribution is  not thermodynamically consistent \cite{vanroijmulderscaling} it enables us to render the orientational averages involved in the elastic moduli analytically tractable as we shall demonstrate in the remainder of this paper.

\section{Elastic moduli of nematic fluids of disks and rods}

\begin{figure}[t]
\begin{center}
\includegraphics[width= \columnwidth]{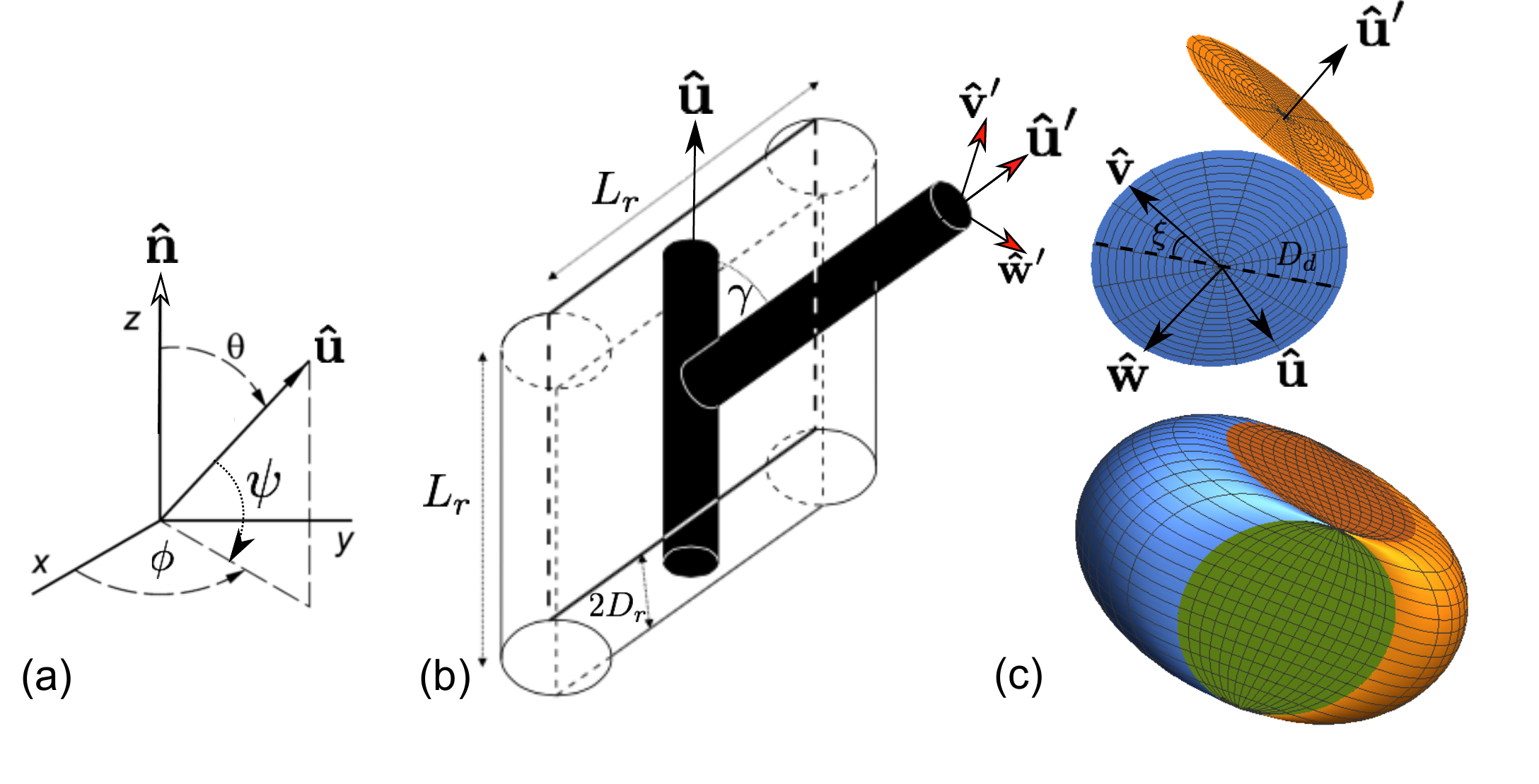}
\caption{ \label{fig2} (a) Overview of the lab frame, nematic director $\bn$, and principal angles used in the present analysis.  (b) Excluded volume between two thin hard rods at fixed mutual angle $\gamma$ and aspect ratio $L_{r}/D_{r} \gg 1$. The orthonormal particle frame $\{ \bhu, \bv, \bw \}$ is indicated by the red arrows. (c) Same for infinitely thin disks with diameter $D_{d}$.  For non-perpendicular orientations ($\gamma \neq \pi/2$) the rods generate a basic lozenge shape while for the disks a slanted sphero-cuboid is obtained.  }
\end{center}
\end{figure}

We begin by presenting an detailed calculation of the splay, twist and bend elastic modes for a simple uniaxial nematic fluid of infinitely thin rigid hard disks of diameter $D$.  The spatial integral required for the elastic moduli takes the form of a generalized or weighted excluded volume:
 \begin{align}
 M_{n}(\bhu, \bhu \p )  &= -\int d \Delta \br  (\Delta \br \cdot \be_{i})^{n} \Phi (\Delta \br, \bhu ,\bhu \p) \nonumber \\
 & = \int d \Delta \br_{dd} (\Delta \br_{dd} \cdot \be_{i} )^{n}, 
 \label{wev}
 \end{align}
 with $n=2$ and $\be_{i} $ ($i=x,y,z$) indicating the Cartesian base vectors. Most importantly,  $\Delta \br_{dd}$ parameterizes the overlap zone between two infinitely flat disks (illustrated in \fig{fig2}) at fixed orientations $\bhu$ and $\bhu \p$ within a particle-based frame $\{ \bw, \bw \p , \bv \} $ with $\bv = (\bhu \times \bhu \p)/|\sin \gamma | $ and $\bw^{(\prime )} = \bhu^{( \prime)} \times \bv$:
 \beq
 \Delta \br_{dd} = -\frac{D_{d}}{2} t_{1} \bw  - \frac{D_{d}}{2} t_{2} \bw\p + \frac{D_{d}}{2} t_{3} [(1-t_{1}^{2})^{\frac{1}{2}} + (1-t_{2}^{2})^{\frac{1}{2}}] \bv,
 \eeq
with integration variables $-1\leq t_{1,2,3} \leq 1$. The Jacobian associated with the transformation from the Cartesian to the particle-based frame reads $J = \frac{D_{d}^{3}}{8} |\sin \gamma |  [(1-t_{1}^{2})^{\frac{1}{2}} + (1-t_{2}^{2})^{\frac{1}{2}}] $. The excluded volume between two hard disks simply follows from \eq{wev} taking $n=0$ and  reads:
\begin{align}
 M_{0}(\bhu, \bhu \p ) &=   \int \Delta \br_{dd} \sim \frac{\pi}{2}D_{d}^{3} | \sin \gamma |.
\end{align}
The kernels  $M_{2}$ needed for the elastic moduli take on a more complicated orientational dependence, namely:
\beq
 M_{2}^{(i)}(\bhu, \bhu \p )  = \frac{\pi D_{d}^{5}}{192}  | \sin \gamma | [22 (\bv \cdot \be_{i})^{2} + 7  (\bw \cdot \be_{i})^{2} + 7  (\bw \p \cdot \be_{i})^{2} ]. 
 \label{m2disk}
\eeq
The remaining task is to perform a double weighted orientational averages in \eq{k123}:
\begin{align}
K_{1}  & \sim \frac{\rho_{d}^{2}}{2} \langle \langle M_{2}^{(y)} (\bhu, \bhu \p )  u_{y} u\p_{y}  \rangle \rangle_{\dot{f}} \nonumber \\
K_{2}  & \sim \frac{\rho_{d}^{2}}{2} \langle \langle M_{2}^{(x)} (\bhu, \bhu \p )  u_{y} u\p_{y}  \rangle \rangle_{\dot{f}} \nonumber \\
K_{3}  & \sim \frac{\rho_{d}^{2}}{2} \langle \langle M_{2}^{(z)} (\bhu, \bhu \p )  u_{y} u\p_{y}  \rangle \rangle_{\dot{f}}. 
\label{moddef}
\end{align}
with $\rho_{d}$ the number concentration of disks. Proceeding toward an explicit calculation of the moduli we consider a simple Gaussian Ansatz \eq{gauss} for the orientational fluctuations around  the colloidal director $\bn$. The derivative needed for the computation of the elastic moduli simply reads $\dot{f}_{G} = \alpha f_{G}$.  In the weak fluctuation limit  $\theta \ll 1$ we write up to leading order  $\bhu \sim \{  \theta \sin \phi , \theta  \cos\phi , 1 \}$. If we further assume uniaxial nematics then $f_{G}$ does not depend on the azimuthal angle $\phi$.  Furthermore, we introduce a dimensionless disk concentration $c_{d} = \rho_{d} D_{d}^{3}$ and express, from here on, all elastic modes in units $k_{B}T/D_{d}$ with $D_{d}$ the cylinder diameter.  

After performing tedious trigonometric manipulations we can express the elastic moduli in terms of the relevant angular variables. Retaining only the leading order terms for small polar angles we obtain a set of rather hefty expressions shown in the Appendix. Fortunately, the results can be greatly compactified by invoking a basic scaling relation for the Gaussian averages, namely $\langle \langle \theta^{n}/|\gamma | \rangle \rangle_{\dot{f}_{G}}  \propto \alpha^{-\frac{n}{2} +\frac{5}{2}} $. Combining this  with the common quadratic dependency of the Gaussian variational parameter with concentration,  $\alpha \sim (4/\pi) (\pi^{2}/16)^{2} c_{d}^{2}$, well-known from bulk nematics \cite{odijkoverview,vroege92}, we readily infer a number of simple scaling expressions for the discotic elastic moduli: 
\begin{align}
K_{1} &\sim 0.303 c_{d}^{3} \nonumber \\
K_{2} &= 2 K_{1}  \nonumber \\
K_{3} &\sim 0.234 c_{d}.
\end{align}
The prefactors were computed numerically even though it may be possible to derive rational prefactors upon extensive mathematical analysis that we did not pursue.  

The large splay-tot-bend ratio $K_{1}/K_{3} \gg1 $ (assuming $c_{d} >1$ in the nematic regime) is in agreement with earlier theoretical predictions \cite{sokalski1982,obrien2011,simonario2016,mulder2018} and experiments \cite{warmerdam1988} on various thermotropic discotic nematics. The interrelation of the different moduli turns out to be quite different from those established for rod-based nematics.  Based on an analogous analysis Odijk \cite{odijkelastic} derived the following results for thin hard rods:
\begin{align}
K_{1} &\sim  \frac{7}{32} c_{r} \nonumber \\
K_{2} & = \frac{1}{3} K_{1}  \nonumber \\
K_{3} &\sim   \frac{\pi}{48} c_{r}^{3}.
\end{align}
where $c_{r} = \rho_{r} L_{r}^{2}D_{r}$  is a dimensionless rod concentration defined in terms of the principal rod dimensions. The results demonstrate that the principal elastic response of discotic nematics stems from the splay ($K_{1}$) and twist ($K_{2}$) modes, while bend elasticity ($K_{3}$) dominates rod-based nematic fluids.  The discrepancy among the modes is expected to be considerable for crowded nematics since all dominant modes scale with the cube of the particle concentration while the minor contributions increase linearly with concentration. The basic interrelation of the elastic moduli for  rod ($N^{r+}$) and disk-based ($N^{d+}$) nematics  are summarized in Table I.

\begin{table*}
\caption{\label{table1}%
Overview of the elastic moduli ratio for nematic (+) and {\em anti}-nematic (-) fluids of hard rods ($r$) and disks ($d$) as predicted from second-virial theory in the limit of asymptotically strong (anti-)nematic order.  The interrelation of the elastic moduli depends on particle concentration $\rho$ or the effective anti-nematic field with strength $W$, specified in the main text.  The field amplitude is tuned by the temperature of the molecular host phase in case of a hybrid colloidal-molecular liquid crystal or by the interspecies interaction in case of a  mixed rod-disk nematic characterized by a ratio of particle dimensions $q=L_{r}/D_{d}$  (see \fig{fig1}). Also shown are the twist-splay and bend-splay ratios that stem from the surface-anchoring of the colloids (indicated by $(s)$). These contributions are only specified for the single component anti-nematic systems.  }
\begin{ruledtabular}
\begin{tabular}{ l c c c c}
\textrm{Phase}&
\textrm{twist-splay: $K_{2}/K_{1}$}& $K^{(s)}_{2}/K^{(s)}_{1}$ &
\textrm{bend-splay: $K_{3}/K_{1}$ } & $K^{(s)}_{3}/K^{(s)}_{1}$ \\
\colrule
$N^{r+}$ rods & 1/3 &   &    $\gg1$  ($\propto \rho_{r}^{2}$) &  \\
$N^{r-}$ rods & 1/3  & 1/3  &     $\ll 1$  ($\propto 1/W$) &  $\ll 1$  ($\propto 1/W$) \\
$N^{d+}$ disks & 2 &    &       $\ll 1$ ($\propto 1/\rho_{d}^{2}$) &   \\ 
$N^{d-}$ disks & 3  & 3  &       $\frac{44}{7}$  &  4 \\
$N^{d+/r-}$ mixed ($q \ll1 $) &  $\approx $ 1 && $\ll $ 1 &\\
$N^{d+/r-}$ mixed ($q \gg1 $)  & $\approx $ 1/3 && $ < $ 1& \\
$N^{r+/d-}$ mixed ($q \ll1 $)  & 3 && $\approx $ 4 &\\
$N^{r+/d-}$ mixed ($q \gg1 $) &  3 && $\gg1 (\propto q ) $ & \\
\end{tabular}
\end{ruledtabular}
\end{table*}

\section{Elastic moduli for anisotropic colloids with surface-anchoring stabilized anti-nematic order}

We now proceed towards analyzing the case of anti-nematic order  in which the particle orientation vectors point perpendicular  to the principal director $\bn = \bz$ without exhibiting a preferred direction of alignment across the $x$-$y$ plane.  

Hybrid colloidal-molecular liquid crystals may be characterized by anti-nematic order that is spontaneously generated through the surface anchoring energy of the anisotropic colloids embedded in a thermotropic liquid crystalline solvent  \cite{Liu2016,Mundoor18}. This is illustrated in \fig{fig1}.  Let us consider a simple Rapini-Papoular  expression for the surface anchoring energy per colloid \cite{rapa}:
\beq
U_{s} = -\frac{1}{2} w_{0}  \int d \bhu \int d{\mathcal S} (\bn \cdot \bn_{0}({\mathcal S}))^{2} f(\bhu \cdot \bn)
\label{rap}
\eeq
with $w_{0} >0$ the surface anchoring coefficient, $\bn$ the uniform director of the molecular host in which the colloid is embedded and $\bn_{0}({\mathcal S})$ the unit vector describing the direction of preferred surface alignment across the colloid surface ${\mathcal S}$. Let us assume that the molecules of the thermotropic background prefer to align parallel to the face of the disk (planar anchoring) or perpendicular to the long axis of the rod (homeotropic anchoring), then we may write in both cases  $\bn_{0}(\xi)  = \cos \xi \bv + \sin \xi \bw $ ($0< \xi < 2 \pi $) in terms of an orthonormal particle frame depicted in \fig{fig2}. It is then easily established that the surface anchoring energy of the rod or disk changes with the polar angle $\theta$ between  the molecular director and the main particle orientation via: 
\beq
U_{s} \sim  - \epsilon_{r/d} \sin ^{2} \theta  
\label{sufac}
\eeq
with $ \epsilon_{r} = \frac{\pi}{4} w_{0} L_{r}D_{r} $ (rods) and $\epsilon_{d} =\frac{\pi}{4} w_{0} D_{d}^{2}$ (disks). Clearly, the surface anchoring energy reaches a  minimum at $\theta = \pi/2$ when the particle orientation is {\em perpendicular} to the molecular director \cite{Liu2016,Mundoor18}. It is important to note that, throughout this work, we assume that the surface anchoring is weak and that nematic distortions generated by the colloidal inclusions are minimal \cite{brochard1970, poulin1997}, a condition that should be satisfied as long as the rod or plate thickness is very small, more specifically $D_{r} \approx L_{d} \ll K/w_{0}$ with $K$ the typical elastic modulus of the molecular host. Taking order-of-magnitude estimates for $K \sim 10^{-12} N$ and $w_{0} \sim 10^{-5} N/m$ one infers  a surface anchoring extrapolation length of about  $K/w_{0} \sim 10^{-1} \mu m $ which is much larger than the typical thickness of most strongly anisotropic mineral colloids  \cite{Mundoor18,davidson-overview}.

Since the basic uniaxial symmetry is retained for anti-nematic order the basic director deformations are identical to those defined in \eq{n123} and we start from the microscopic definitions laid out in  Sec. II to analyze the nemato-elastic properties.

\subsection{Rod-based anti-nematics}

Starting with hard rods,  we define $M_{2}$ as the generalized excluded volume kernel for two hard cylinders with aspect ratio tending to infinity ($L_{r}/D_{r} \rightarrow \infty$) \cite{straley76, wensinkjackson}:
\begin{align}
M_{2}^{(i)} (\bhu, \bhu \p) & =  -\int d \Delta \br  ({\bf \Delta r} \cdot \be_{i}) ^{2} \Phi (\Delta \br, \bhu ,\bhu \p) \nonumber \\
& \sim \frac{1}{6} L_{r}^{4}D_{r} | \sin \gamma | (u_{i}^{2} + (u_{i} \p)^{2})
\end{align}
It is easily checked that the zeroth order moment simply yields the excluded volume $M_{0} = 2L_{r}^{2}D_{r} | \sin \gamma | $ of the lozenge depicted in \fig{fig2}.

Defining a meridional angle $\psi = \frac{\pi}{2} - \theta$, parameterizing the unit vector $\bhu = \{ \cos \psi \sin \phi, \cos \psi \cos \phi , \sin \psi\} $ and taking the leading order contributions for small $\psi \ll 1$ in the orientational kernels above we obtain explicitly:
\begin{align}
K_{1}  & \sim \frac{c_{r}^{2}}{32} \langle \langle    | \sin \gamma | \cos \Delta \phi \cos \psi \cos \psi \p (\cos^{2} \psi + \cos ^{2} \psi \p)  \rangle \rangle_{\dot{f}_{U}} \nonumber \\
K_{2} & \sim   \frac{1}{3} K_{1}  \nonumber \\
K_{3} & \sim  \frac{c_{r}^{2}}{24}    \langle \langle     | \sin \gamma | \cos \Delta \phi \cos \psi \cos \psi \p (\sin^{2} \psi + \sin ^{2} \psi \p)    \rangle \rangle_{\dot{f}_{U}} 
\label{kdef}
\end{align}
with $ \sin^{2} \gamma  =  1- (\cos \Delta \phi \cos \psi \cos \psi \p + \sin \psi \sin \psi \p)^{2}$.  The leading order contribution for small meridional angles can be obtained by expand the trigonometric functions up to second order in $\psi$, using circular coordinates $\psi = R \sin \chi $ and $\psi \p = R \cos \chi$. The integration over the relative azimuthal angle can then be cast in terms of complete elliptic integrals $E(1-R^{2})$ of which we retain only the dominating contribution for $R \ll 1$. The mathematical details of this procedure are outlined in Ref. \cite{wensinkrodplate2001}.  After some algebra, we  obtain up to second order in $R$:
\begin{align}
K_{1}  & \sim \frac{c_{r}^{2}}{16 \pi}  \langle \langle   R^{2} \sin \chi \cos \chi  (4\ln 2 - 2 -2 \ln R) \rangle \rangle_{\dot{f}_{U}} \nonumber \\
K_{2} & \sim   \frac{1}{3} K_{1}  \nonumber \\
K_{3} & \sim \frac{c_{r}^{2}}{24 \pi}   \langle \langle   R^{4} \sin \chi \cos \chi  (4\ln 2 - 2 -2 \ln R)  \rangle \rangle_{\dot{f}_{U}} 
\label{kanti}
\end{align}
The splay-twist ratio $K_{1}/K_{2} =3$ is identical to the one predicted for conventional rod-based nematics, while the bend-splay ratio $K_{3}/K_{1} $ turns out much smaller than unity, similar to a that of a discotic nematic.

Let us now attempt an explicit calculation of the averages by taking the anti-nematic orientation distribution function for the rods proposed by Mundoor {\em et al.}  \cite{Mundoor18} to describe the anti-nematic order of rods in a hybrid molecular-colloidal liquid crystal:
\beq
f_{U}(\theta) = {\mathcal N} \exp \left [ -W \pp (\cos \theta) \right ] 
\label{fu}
\eeq
in terms of a variational parameter $W$ quantifying the degree of anti-nematic order:
\beq
-W =  \frac{5 \pi}{16} c_{r} S_{r} - \epsilon_{r}.  
\eeq
and $ {\mathcal N} $ a normalization constant. Furthermore, $S_{r} <0 $ is the nematic order parameter being negative for the anti-nematic state. Crucially, $\epsilon_{r} >0$ is the strength of a temperature-controlled effective external field imparted by the surface anchoring properties of the rods with the molecular host forcing the rods to align perpendicular to the molecular director (cf. \eq{sufac}). 

In the experimental system considered in Ref. \cite{Mundoor18}, the effective field originating from the surface-anchoring  energy is quite strong ($\epsilon_{r} \gg 1$ and $W \gg  1$) and so will be the degree of anti-nematic order.  The rod vectors will therefore adopt a very small equatorial tilt angle $\psi = \frac{\pi}{2} - \theta \ll 1$ which justifies the use of a simple Gaussian asymptotic of the form:
\begin{align}
f_{U}(\psi ) & \sim \left ( \frac{3W}{2 \pi} \right )^{1/2} \exp \left  (  -\frac{3}{2} W \psi^{2} \right ), \hspace{0.5cm} W \gg 1 \nonumber \\ 
\dot{f}_{U}(\psi) & \sim \frac{\partial f_{U}}{\partial \psi} \sim -3 W \psi f_{U} (\psi) 
\label{fudot}
\end{align}
with the variational parameter $W$ related to the nematic order parameter via $S_{r} \sim (1-W)/2 W$.
The limit $S_{r} \rightarrow -\frac{1}{2}$ for $W \rightarrow \infty$ is easily verified.  The double orientational averages in \eq{kanti} are given by:
\beq
 \langle \langle (\cdot) \rangle \rangle_{\dot{f}_{U}} =  \frac{27W^{3}}{2 \pi}  \int_{0}^{2 \pi} d \chi  \sin \chi \cos \chi \int_{0}^{\infty} dR R^{3} ( \cdot ) e^{-\frac{3}{2} W R^{2}} 
\label{dangle}
\eeq
and the remaining integrals can be solved in closed form with the help of the mathematical analysis reported in Ref. \cite{wensinkrodplate2001}. Rearranging terms we obtain for the elastic moduli of an anti-nematic phase of rods: 
\begin{align}
K_{1} &   \sim \frac{ 1}{ 16\pi } c_{r}^{2} (\ln W + C_{1}) \nonumber \\ 
K_{2} & \sim   \frac{1}{3} K_{1}  \nonumber \\
K_{3} & \sim  \frac{1}{ 12 \pi} c_{r}^{2} \frac{ \ln W + C_{3}}{W}
\end{align}
with constants $C_{1} = \gamma_{E} -7/2  + \ln 24 \approx 0.255269$ and $C_{3} =\gamma_{E} -23/6 + \ln 24  \approx -0.0780638$ and $\gamma_{E}$ being Euler's constant. In good approximation we may assume that $S_{r} \approx -\frac{1}{2}$ so that $W \sim \epsilon_{r} + \frac{5 \pi}{32} c_{r}$.  
This renders the elastic moduli fully explicit in terms of the rod concentration $c_{r}$ and field strength $\epsilon_{r} \gg 1$. 

\subsection{Discotic anti-nematics}

We may repeat the analysis for an anti-nematic arrangement of hard disks with their normals pointing perpendicular to the nematic director. Starting from the kernel \eq{m2disk}, we insert the anti-nematic parameterization of the unit vectors of the particle frame, introduced above for the rod case, and expand up to leading order. The anti-nematic moduli for disks take on a simple form and turn out to differ only by a  constant factor:
\begin{align}
K_{1}  & \sim  \frac{7 \pi c_{d}}{1536}  \langle \langle \frac{\cos \Delta \phi (\sin \Delta \phi)^{2}}{|  \gamma |} \rangle \rangle_{\dot{f}_{U}} \nonumber \\
K_{2} & \sim    3 K_{1} \nonumber \\
K_{3} & \sim  \frac{44}{7} K_{1}
\label{kantidisk}
\end{align}
The angular average in the first expression can be worked out in analytical form using the coordinate transform introduced below \eq{kdef}  which enables us to cast the average over the azimuthal angle $\Delta \phi$ in terms of elliptic integrals for which the leading contributions for small $R$ are of logarithmic form. This leads to:
\beq
K_{1}  \sim  \frac{7 \pi c_{d}^{2}}{768}    \langle \langle R^{2} \sin \chi \cos \chi  (4 \ln 2 -4 - 2 \ln R) \rangle \rangle_{\dot{f}_{U}} 
\eeq
Employing the double averaging \eq{dangle} weighted by the orientational distribution \eq{fu}, with $W$ suitably redefined in terms of the order parameter and concentration for the disks
$ -W  = \frac{5 \pi^{2}}{32} c_{d} S_{d} - \epsilon_{d} $
where $S_{d} >0$. Working out the averages we find that the splay modulus exhibits  a similar logarithmic scaling with field strength $W$ as observed for the rods, namely
\beq
K_{1} \sim   \frac{7 \pi^{3} c_{d}^{2}}{768} (\ln W + C_{1})
\eeq
with $C_{1} = \gamma_{E} - 11/2  + \ln 24$. 

The main characteristics of the predicted elastic modes  gathered so far are summarized  in Table I.
The logarithmic form of the moduli implies a much weaker concentration (or temperature) sensitivity for anti-nematic systems than expected for conventional nematics. 

Also, while both splay and twist elasticity of anti-nematically ordered rods  are (logarithmically) increasing functions of $W$, the bend contribution $K_{3} \propto  W^{-1} \ln W$ exhibits a maximum at $W=e\approx 2.71$ and drops at large $W$ as the degree of anti-nematic order increases. The opposite trend is observed for the disks,  where the bend elastic term  increases monotonically with field strength $W$.  Clearly, the bend-splay ratio of nematic materials can be boosted considerably by introducing anti-nematically ordered colloidal disks.

\section{Surface-anchoring elasticity for hybrid colloidal-molecular nematics} 

\begin{figure}[t]
\begin{center}
\includegraphics[width= \columnwidth]{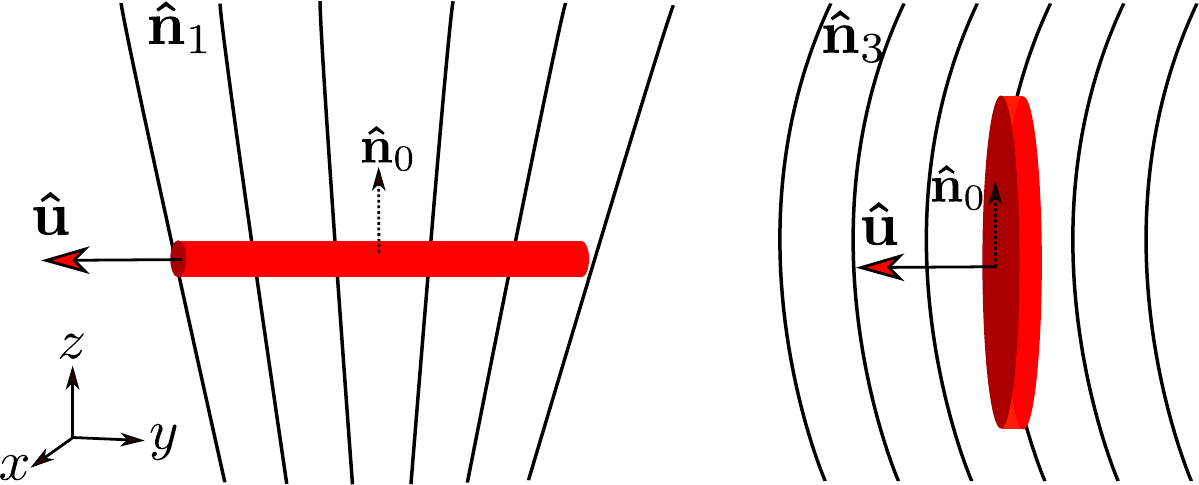}
\caption{ \label{fig1a}  A splay deformation of a thermotropic liquid crystal is suppressed by the presence of a thin rigid rod with homeotropic surface anchoring oriented perpendicular to the average nematic director, while a thin disk with planar surface anchoring counteracts a bend fluctuation of the host director. The surface-anchoring effect will be proportional to the colloid size (rod length or disk diameter).  The direction of preferred surface alignment is indicated by the vector $\bn_{0}$. The director fluctuations $\bn_{1,2,3}$ have been specified in \eq{n123}.}
\end{center}
\end{figure}

So far, our attention has been on describing anti-nematic elasticity transmitted through colloid-colloid interactions. However, in case of anisotropic colloidal particles immersed in a thermotropic liquid crystals the inclusion of even a single colloid will perturb the elastic properties of the embedding thermotropic liquid crystals. The mechanism is quite simple; a long-wavelength perturbation of the molecular director in the presence of a colloidal particle will locally alter the anchoring condition at the particle surface and induce a change in free energy. This is illustrated in \fig{fig1a}.  The associated free energy change will be proportional to the main size of the colloidal inclusion and can be identified with a surface-anchoring mediated elasticity that we will quantify for both rods and platelets. 

\subsection{Rods with homeotropic surface anchoring}

Starting with infinitely thin rods with homeotropic surface anchoring, the molecular director is preferentally aligned along the surface normal vector $\bn_{0}  = \cos \xi \bv  + \sin \xi \bw  $  the rod surface in terms of an orthonormal particle frame $\{ \bhu, \bv , \bw \}$ (see \fig{fig2}). The Rapini-Papoular expression \eq{rap} can be easily  generalized to the case of a non-uniform molecular director field $\bn({\bf R})$ formulated in \eq{n123}. Expanding the surface anchoring free energy up to  ${\mathcal O} (\epsilon^{2})$ we can deduce the free energy change associated with infinitely weak long-wavelength molecular director deformations. The associated elastic contributions originating from the surface anchoring condition are then defined as $K_{n}^{(s)} = \delta F_{s}/ \frac{1}{2} \epsilon^{2} V$ which read in explicit form: 
\begin{align}
K_{n}^{(s)} & = -\frac{1}{4} w_{0} \rho_{r}  D_{r} \int d \bhu  \int_{0}^{2 \pi} d \xi  \int_{-L_{r}/2}^{L_{r}/2} dt  \left ( t \bhu \cdot \be_{n} \right )^{2} \nonumber \\ 
& \times \left [ (\by \cdot \bn_{0}(\xi) )^{2} - (\bz \cdot \bn_{0}(\xi) )^{2}\right ] f(\bhu \cdot \bn)
\label{kanchor}
\end{align}
with $\be_{1} = \by$, $\be_{2} = \bx$ and $\be_{3} = \bz$ as per the different modes [cf. \eq{n123}]. The expression above involves an integral over the rod contour $t$ probing the protrusion of the rod in the direction  $\be_{n}$ along which the host director is distorted.  The integrals are easily solved using the Gaussian orientational distribution for the anti-nematic configuration formulated above.  The resulting expressions for the elastic modes take a particularly simple form:
\begin{align}
K_{1}^{(s)} & \sim \frac{\pi}{128} w_{0}  c_{r} \frac{L_{r}}{D_{r}} \nonumber \\ 
K_{2}^{(s)} & =   \frac{1}{3} K_{1}^{(s)}  \nonumber \\
K_{3}^{(s)} & \sim   \frac{\pi}{48} w_{0}  c_{r}\frac{L_{r}}{D_{r}}\frac{1} {W}
\label{ks}
\end{align}
Similar to rod-interaction driven elasticity the twist contribution is one third of the splay term, and that splay elasticity dominates the other two modes (since $W>0$). An important distinction, however, is that the surface anchoring-mediated contributions, representing principally a single rod effect, increase linearly with rod concentration. Clearly, the surface-anchoring elasticity is most noticeable for the splay mode while practically negligible for the bend mode provided the rods are strongly anti-nematic ($W \gg 1$).

\subsection{Disks with planar surface anchoring}

Turning now to the case of infinitely thin disks with planar surface anchoring we formulate the surface-anchoring elasticity analogous to \eq{kanchor}:
\begin{align}
K_{n}^{(s)} & = -\frac{1}{2} w_{0} \rho_{d}   \int d \bhu  \int_{0}^{2 \pi} d \xi  \int_{0}^{D_{d}/2} dr r \oint d {\bf \hat{r}} \left ( r{\bf \hat{r}}  \cdot \be_{n} \right )^{2} \nonumber \\ 
& \times \left [ (\by \cdot \bn_{0}(\xi) )^{2} - (\bz \cdot \bn_{0}(\xi) )^{2}\right ] f(\bhu \cdot \bn)
\label{kdiskanchor}
\end{align}
with ${\bf \hat{r}} = \cos \xi \bv + \sin \xi  \bw $ parameterizing the disk surface (see \fig{fig2}). The resulting expressions (in units ) are again quite simple and read:
\begin{align}
K_{1}^{(s)} & \sim \frac{\pi^{2}}{1024}  w_{0}  c_{d}  \nonumber \\ 
K_{2}^{(s)} & =   3 K_{1}^{(s)}  \nonumber \\
K_{3}^{(s)} & = 4 K_{1}^{(s)}
\label{ksdisk}
\end{align}
Since disks are much more isotropic objects than slender rods they tend protrude in all Cartesian directions. Consequently, their surface anchoring properties exert a considerable impact on {\em all} director deformations of the molecular host structure with the bend mode being most affected.  

For either colloid shape, the importance of the correlation-driven versus surface-anchoring mediated elasticity depends quite sensitively on the anchoring strength $w_{0}$,  the main particle dimension and the concentration of inclusions which should be restricted to the regime $c <1$ to keep the colloids from aligning in-plane and preserve anti-nematicity.  
We anticipate that for hybrid nematics with strongly nematic colloidal order  ($S>0$ and $c \gg 1$) the correlation-mediated elasticity will outweigh the one driven by surface anchoring effects.

\section{Elasticity generated by interspecies interactions in rod-disk nematics} 

We conclude our analysis by investigating the elastic contributions arising from rod-disk interactions that characterize the mixed colloidal nematics illustrated in \fig{fig1}b. 
In these situations, the nemato-elasticity is primarily transmitted by the principal component which is assumed to be ordered nematically, with the anti-nematic dopant perturbing the elastic properties of the mixed phase through rod-disk cross-interactions that we will attempt to quantify in the following.

In keeping with our original model  of infinitely thin hard uniaxial cylinders we identify the principal dimensions as the disk diameter $D_{d}$ and rod length $L_{r}$,  and parameterize the rod-disk ($rd$) excluded volume as follows:
\beq
\Delta \br_{rd} = \frac{D_{d}}{2} t_{1} \sin \xi \bv + \frac{D_{d}}{2} t_{1}  \cos \xi \bw + \frac{L_{r}}{2} t_{2} \bhu \p
\eeq
 in terms of a particle-based orthogonal frame $\{ \bhu, \bv, \bw \}$ for the disk  and $\{ \bhu \p, \bv \p , \bw \p \}$ for the rod, with integration variables $0< t_{1,2} <1$ and $0< \xi < 2 \pi$. The Jacobian associated with the coordinate transformation reads $J = \frac{1}{8} L_{r}^{2}D_{d} t_{1} |\bv \cdot (\bw\times \bhu \p) |  $ so that:
 \begin{align}
M_{0}(\gamma) &=  \int d \Delta \br_{rd} \sim  \frac{\pi}{4} L_{r}D_{d}^{2} | \cos \gamma | 
 \end{align}
 yields the excluded volume between a rod and a disk with orientation vectors making an angle $\gamma$. As in the preceding calculations we have neglected all subleading correction terms arising from the particles' finite thicknesses which are notoriously difficult to quantify \cite{vroijthesis} but are deemed unimportant for sufficiently slender mineral colloids. The second-order kernel reads:
 \beq
M_{2}^{(i)}(\bhu, \bhu \p) \sim \frac{\pi}{192} D_{d}^{5} | \cos \gamma |  \left [ 4 q^{3} (u^{\prime}_{i})^{2} + 3 q (v_{i}^{2} + w_{i}^{2}) \right ], 
 \eeq
(with $i=x,y,z$) in terms of the ratio of the principal particle dimensions $q = L_{r}/D_{d}$. As before, the moduli are given by [cf. \eq{moddef}]:
 \begin{align}
K_{1} & \sim \frac{\rho_{r}\rho_{d}}{2} \langle \langle  M^{(y)}_{2} (\bhu, \bhu \p) u_{y} u\p_{y}  \rangle \rangle_{\dot{f}_{U}} \nonumber \\
K_{2} & \sim \frac{\rho_{r}\rho_{d}}{2} \langle \langle  M^{(x)}_{2} (\bhu, \bhu \p)  u_{y} u\p_{y}  \rangle \rangle_{\dot{f}_{U}} \nonumber \\
K_{3} & \sim  \frac{\rho_{r}\rho_{d}}{2} \langle \langle  M^{(z)}_{2} (\bhu, \bhu \p)  u_{y} u\p_{y} \rangle \rangle_{\dot{f}_{U}} 
\end{align}
in terms of the partial concentrations $\rho_{r}$ and $\rho_{d}$. It is important to note that both components obey distinctly different orientation distributions where one species is ordered nematically with the other adopting an anti-nematic configuration, so that the double orientational averaging proceeds via unequal orientational distributions, i.e. $\dot{f}_{U}(\bhu) \neq \dot{f}_{U}(\bhu \p)$.  In the following we shall demonstrate that these orientational averages do not need be computed in explicit form in order to gauge some basic scaling properties pertaining to the ratio of the elastic moduli.

The first system under consideration is labeled $N^{d+/r-}$ and consists of a nematic arrangement of disks mixed with an anti-nematic organization of rods (see \fig{fig1}).  As in the preceding, we use $k_{B}T/D_{d}$ as a force unit to render the scaling expressions for the elastic moduli dimensionless and consider normalized colloid concentrations $c_{r}$ and $c_{d}$. The results  are as follows:
 \begin{align}
K_{1} & \sim  \frac{c_{r}c_{d}}{2} \langle \langle   \frac{\pi}{128} \theta \cos \Delta \phi | \cos \gamma |  \left ( \frac{q^{-1}}{\sin^{2} \gamma} + q \right )  \rangle \rangle_{\dot{f}_{U}} \nonumber \\
K_{2} & \sim  \frac{c_{r}c_{d}}{2} \langle \langle    \frac{\pi}{128} \theta \cos \Delta \phi | \cos \gamma |  \left ( \frac{q^{-1}}{\sin^{2} \gamma} + \frac{q}{3} \right )  \rangle \rangle_{\dot{f}_{U}} \nonumber \\
K_{3} & \sim   \frac{c_{r}c_{d}}{2} \langle \langle \frac{\pi}{96} q \theta (\psi \p)^{2} \cos \Delta \phi | \cos \gamma |  \rangle \rangle_{\dot{f}_{U}} 
\end{align}
 with $\theta \ll 1 $ the polar angle of the disks and $\psi \p \ll 1$ the meridional one for the rods (see \fig{fig2}). Furthermore, we note that $|\cos \gamma| \ll 1$ and $\sin^{2} \gamma \sim {\mathcal O}(1)$.
 
The second case,  $N^{r+/d-}$,  is a rod-based  nematic in which the disks are dispersed anti-nematically. In this situation the moduli take on the following form:
 \begin{align}
K_{1} & \sim  \frac{c_{r}c_{d}}{2} \langle \langle   \frac{\pi}{512} \theta \p \cos \Delta \phi | \cos \gamma |   \frac{q^{-1}}{\sin^{2} \gamma}  \rangle \rangle_{\dot{f}_{U}} \nonumber \\
K_{2} & = 3 K_{1} \nonumber \\
K_{3} & \sim   \frac{c_{r}c_{d}}{2} \langle \langle \frac{\pi}{128}  \theta\p  \cos \Delta \phi | \cos \gamma | \left ( \frac{q^{-1}}{\sin^{2} \gamma} + 2 q \right )   \rangle \rangle_{\dot{f}_{U}} 
\end{align}  
Refraining from any further analysis, we easily infer the interrelation of the elastic moduli for the two mixed rod-disk uniaxial nematics depicted in \fig{fig1}b. The results obtained so far are tabulated in Table I. It is apparent that the elasticity of the mixed nematics can be carefully tuned by varying the ratio $q$ of the rod length versus disk diameter, as well as by changing the partial concentrations.  A general conclusion one could draw from looking at the results for the mixed systems is that  anti-nematically order disks  have a much more significant impact on the elastic properties (in particular enhancing the bend mode) of rod-dominated nematics ($N^{r+/d-}$) while rod-shaped inclusions turn out to only weakly influence the elasticity of a discotic nematic.    Furthermore, the results above suggest an interesting non-monotonic trend with the  rod-disk size ratio $q$. This happens for the splay and twist elasticity of the discotic nematic and the bend elasticity of the rod-based nematic, where the moduli exhibit a minimum around $q \sim 1$, that is, for rods and disks of about equal particle dimensions.

We wish to demonstrate that it is possible to specify the effective anti-nematic field strength of the rod-disk nematics in terms of relevant component variables by considering the coupled rod- and disk orientation distributions for a mixed uniaxial nematic proposed in Ref. \cite{lekkerstro1984}:
\begin{align}
f_{r}(\cos \theta)  & \sim \exp \left [ \frac{5}{4}( \rho_{r} v_{ex}^{rr}S_{r} - 2 \rho_{d}  v_{ex}^{rd}S_{d} ) \pp (\cos \theta)\right ] \nonumber \\
f_{d}(\cos \theta) & \sim \exp \left [ \frac{5}{4}( -2 \rho_{r} v_{ex}^{rr}S_{r} + \rho_{d}  v_{ex}^{rd}S_{d} ) \pp (\cos \theta)\right ] 
 \end{align}
in terms of the isotropized excluded volume of two rods $v_{ex}^{rr} = (\pi/4) L_{r}^{2}D_{r}$ and a rod-disk pair $v_{ex}^{rd} = (\pi/8) L_{r}D_{d}^{2}$. The order parameters are such that $S_{r} >1$ and $S_{d} <0$ for the rod-dominated nematic ($N^{r+/d-}$) and $S_{d} >1$ and $S_{r} <0$ for the disk-dominated one ($N^{d+/r-}$).  Comparing with \eq{fu}) we immediately deduce that the effective anti-nematic field strength for the two mixed nematics in \fig{fig1} read:
 \begin{align}
 \epsilon_{r} &\sim \frac{5 \pi}{16} c_{d} q  S_{d}, \qquad \qquad (N^{d+/r-})  \nonumber \\
 \epsilon_{d} &\sim \frac{5 \pi}{16} c_{r} q^{-2} \ell_{r} S_{r}, \qquad  (N^{r+/d-}) 
  \end{align}
with $\ell_{r}= L_{r}/D_{r} \gg 1$ the length-to-width (aspect) ratio of the rods.  We observe that, in line with expectation,  the effective anti-nematic field imposed on one species imparted by the nematic alignment of the other is proportional to the degree of nematic order (note that the values of $S_{r,d}$ are positive and close to unity) and the concentrations $c_{r,d} \gg 1$ of the other component. 

Keeping in mind that validity of our asymptotic analysis requires that the anti-nematic field strength be much larger than the thermal energy $k_{B}T$ we infer that the condition $\epsilon_{r,d} \gg 1$ imposes two important criteria  on the size disparity of the components, namely $q > 1$ and  $ q^{-2} \ell_{r} >1 $. The latter criterion is easily satisfied for rods in the Onsager limit $\ell_{r} \rightarrow \infty$ whereas the former imposes the rod length to be at least the disk diameter. 

A much stricter criterion that we have not addressed here is the thermodynamic stability of the uniaxial mixed nematic with respect to the formation of a biaxial phase in which the  rods and disks are each aligned along mutually orthogonal directors. Intuitively, for rod-disk mixtures, one would expect uniaxial order to be preferable over biaxial nematics  for strongly {\em asymmetric} mixtures with a strong disparity in the excluded volume among the components \cite{roijmulder1994,wensinkrodplate2001,varga2002}, which is broadly satisfied through the aforementioned criteria. In most practical situations studied thus far, the biaxial nematic tends to be unstable with respect to demixing into two fractionated uniaxial nematics \cite{camp1997, kooij2000,vroege2014}.

\section{Conclusion}

Inspired by recent experimental advances in the fabrication of well-controlled composite nematics \cite{Mundoor18,lagerwall2012} involving a colloidal component with distinct {\em anti-nematic} order  we have embarked on computing the elastic moduli associated with such anti-nematically ordered anisotropic particles. For simplicity we have kept our focus on simplified but emblematic models for lyotropics, namely  hard cylindrical rods and hard disks with vanishing particle thickness and address the elastic properties through Onsager's second-virial theory, suitably extended by Straley \cite{straley76} to capture the effect of weak deformations of the nematic director field.  
While the elastic moduli for common nematics are in full agreement with those reported in experimental, theoretical and simulation studies, the elastic moduli for the anti-nematic case had not yet been identified and unveil a remarkable logarithmic scaling with the degree of anti-nematic order. More importantly,   the interrelation between splay, twist and bend moduli turns out to be quite different from that of conventional nematics composed of colloidal rods. 

Most notably, we find that the bend elasticity of rods vanishes with increasing anti-nematic order while the bend modulus for disks increases as the disks adopt a more pronounced anti-nematic configuration.  In addition to the elasticity imparted by colloidal interactions, we quantify the effects of colloid surface anchoring which alters the deformation energy of the director field of the molecular liquid crystal in which the colloids are immersed. These moduli are all linear in colloid concentration and increase with the colloid size and the surface anchoring amplitude. We find that the ratio of the twist-splay and bend-splay elastic modes induced by surface anchoring effects are qualitatively similar to those generated by colloid interactions.

Extending our treatment to mixed uniaxial rod-disk nematics we present a preliminary analysis of the elastic moduli for coupled rod-disk interactions and argue that bend fluctuations of rod-dominated  nematics can be strongly suppressed by adding disk-shaped inclusions.  
A full analytical characterization of the elastic moduli for mixed rod-disk nematics in terms of the partial concentrations of the components, not attempted in this paper,  seems a highly non-trivial task but should at least be feasible numerically using the preliminary scaling expressions for the moduli and the Gaussian orientational distributions proposed in this work. 

It would be intriguing to compare our predictions with experimental measurements of the moduli for e.g. colloidal rod-disk mixtures or other composite nematic phases involving anti-nematic order of one or several  components.  For thermotropics, there are a number of papers reporting experimental observations of remarkable changes in elastic response upon adding colloidal dopants, often (but not always) leading to a strong reduction of the elastic resistance of the composite material \cite{sridevi2010, ghosh2011,lapanik2016,madhuri2016}.  To the best of our knowledge no such experimental data  have been reported thus far for the lyotropic systems under scrutiny and we hope that our work will stimulate experimental and simulation efforts to characterize the intricate nemato-elasticity in mixed colloidal or hybrid molecular-colloidal nematics. We finally wish to point out that it should be feasible to extend the current second-virial analysis for biaxial colloidal composites for which there are twelve independent elastic moduli \cite{brand1982,govers1984}. It would be intriguing to theoretically quantify those moduli along the lines of the present analysis and compare with experimental measurements in biaxial hybrid molecular-colloidal materials. Efforts in this direction are currently being undertaken.

\section*{Acknowledgement}

I am grateful to Ivan Smalyukh  of the University of Colorado  (Boulder, USA) for fruitful discussions and for stimulating my interest in this topic. 

\section*{Appendix}
We show here the explicit angular dependence of the discotic elastic constants: 
\begin{widetext}
\begin{align}
K_{1} &\sim -\frac{c_{d}^{2}}{2} \langle \langle  -\frac{7 \pi}{192} \frac{\theta^{2} \theta^{\prime 2} }{| \gamma |}  -\frac{3 \pi}{64} \frac{\theta^{2} \theta^{\prime 2}  \cos 2 \Delta \phi}{|  \gamma |} + \frac{\pi}{24} \frac{(\theta^{3} \theta \p + \theta \theta ^{\prime 3}) \cos \Delta \phi}{|  \gamma |} \rangle \rangle _{\dot{f}_{G}} \nonumber \\
K_{2} &\sim -\frac{c_{d}^{2}}{2} \langle \langle   -\frac{11 \pi}{192} \frac{\theta^{2} \theta^{\prime 2} }{|  \gamma |}  -\frac{3 \pi}{64} \frac{\theta^{2} \theta^{\prime 2}  \cos 2 \Delta \phi}{|  \gamma |} + \frac{5 \pi}{96} \frac{(\theta^{3} \theta \p + \theta \theta ^{\prime 3}) \cos \Delta \phi}{|  \gamma |}\rangle \rangle _{\dot{f}_{G}} \nonumber \\
K_{3} &\sim -\frac{c_{d}^{2}}{2} \langle \langle   \frac{7 \pi}{384} \frac{(\theta^{5} \theta^{\prime} + \theta \theta^{\prime 5} ) \cos \Delta \phi}{|  \gamma |} -  \frac{7 \pi}{192} \frac{(\theta^{2} \theta^{\prime  4} + \theta^{4} \theta^{\prime 2} ) (\cos \Delta \phi)^{2}}{|  \gamma |} + \frac{3 \pi}{64} \frac{\theta^{3} \theta^{\prime  3} \cos \Delta \phi }{|  \gamma |}- \frac{\pi}{96} \frac{\theta^{3} \theta^{\prime  3} \cos \Delta \phi \cos 2 \Delta \phi}{|  \gamma |} \rangle \rangle _{\dot{f}_{G}}. 
\end{align}
\end{widetext}
with $\theta$ and $\theta^{\prime}$ the polar angles, $\Delta \phi = \phi - \phi \p$ the relative azimuthal angle and $\gamma$ the angle between the orientation vectors of the disks (see  \fig{fig2}).



\bibliographystyle{apsrev}
\bibliography{refs}

\end{document}